\documentclass[aps,prd,twocolumn,superscriptaddress,tightenlines,nofootinbib]{revtex4-1}
\usepackage[T1]{fontenc}
\usepackage{microtype}
\usepackage[textsize=scriptsize,backgroundcolor=red!70,linecolor=red]{todonotes}
\usepackage{booktabs}
\usepackage{dcolumn}
\usepackage{amsmath}
\usepackage{amsfonts}
\usepackage{amssymb}
\usepackage{graphicx}
\usepackage{ulem}
\usepackage{hyperref}
\usepackage[all]{hypcap}
\usepackage{paralist}
\usepackage{multirow}

\usepackage{graphicx}
\usepackage{dcolumn}
\usepackage{bm}
\usepackage{epsf}
\usepackage{dcolumn}
\def\be{\begin{equation}}
\def\ee{\end{equation}}
\def\bea{\begin{eqnarray}}
\def\eea{\end{eqnarray}}
\def\gsim{\ \rlap{\raise 2pt\hbox{$>$}}{\lower 2pt \hbox{$\sim$}}\ }
\def\lsim{\ \rlap{\raise 2pt\hbox{$<$}}{\lower 2pt \hbox{$\sim$}}\ }
\def\dslash{\kern-4pt \not{\hbox{\kern-2pt $\partial$}}}
\def\pslash{\not{\hbox{\kern-2pt p}}}

\def\beq{\begin{equation}}
\def\eeq{\end{equation}}

\begin{document}
\DeclareGraphicsExtensions{.eps,.ps}


\title{Implications of the first evidence for coherent elastic scattering of reactor neutrinos}



\author{Jiajun Liao}
\email[Email Address: ]{liaojiajun@mail.sysu.edu.cn}
\affiliation{School of Physics, Sun Yat-sen University, Guangzhou, 510275, China}

\author{Hongkai Liu}
\email[Email Address: ]{lliu.hongkai@campus.technion.ac.il}
\affiliation{Department of Physics, Technion – Israel Institute of Technology, Haifa 3200003, Israel}
 
\author{Danny Marfatia}
\email[Email Address: ]{dmarf8@hawaii.edu}
\affiliation{Department of Physics and Astronomy, University of Hawaii at Manoa, Honolulu, HI 96822, USA}

\begin{abstract}
 
The recent evidence for coherent elastic neutrino-nucleus scattering (CE$\nu$NS) in the NCC-1701 germanium detector using antineutrinos from the Dresden-II nuclear reactor is in good agreement
with standard model expectations.  However, 
we show that a $2\sigma$ improvement in the fit to the data can be achieved if the quenching factor is described by a modified Lindhard model. 
We also place constraints on the parameter space of a light vector or scalar mediator that couples to neutrinos and quarks, and on a neutrino magnetic moment. We demonstrate that the constraints are quite sensitive to the quenching factor at low recoil energies by comparing constraints for the standard Lindhard model with those by marginalizing over the two parameters of the modified Lindhard model.

\end{abstract}
\maketitle

{\bf Introduction.} 
Coherent elastic neutrino-nucleus scattering (CE$\nu$NS) is a process in which low-energy neutrinos  scatter off the entire nucleus~\cite{Freedman:1973yd}. This process was 
first observed by the COHERENT collaboration in 2017 using a pion-decay-at-rest ($\pi$DAR) neutrino source with a cesium-iodide detector~\cite{Akimov:2017ade}, and later confirmed with an argon detector at more than 3$\sigma$ C.L. with the same source~\cite{Akimov:2020pdx}. The observation of CE$\nu$NS is a milestone in neutrino physics, and opens a new window to probe neutrino and nuclear physics at low energies~\cite{review}. 

Nuclear power reactors are attractive as antineutrino sources for CE$\nu$NS experiments because they provide very high neutrino fluxes. However, because reactor neutrinos have lower energies and larger backgrounds compared to $\pi$DAR sources, observing CE$\nu$NS with reactor antineutrinos is difficult. The CONNIE~\cite{Aguilar-Arevalo:2019jlr} and CONUS~\cite{Bonet:2020awv} experiments have managed to place constraints on CE$\nu$NS with reactor neutrinos using a silicon and germanium detector, respectively. 

Building on earlier work~\cite{Colaresi:2021kus}, recently, a first hint of CE$\nu$NS using reactor neutrinos has been reported in Ref.~\cite{NCC}. A low-noise 3~kg p-type point contact germanium detector (named NCC-1701) was placed at a distance of {\color{red}$\sim10$}~m from the 2.96 GW Dresden-II  power reactor for a 96.4 day exposure.  The evidence is supplemented by a new measurement of the germanium quenching factor~\cite{Collar:2021fcl} and better energy resolution. 
In this Letter, we analyze data from NCC-1701 to study their implications for the quenching factor and new physics at energies not yet probed by CE$\nu$NS. 

{\bf CE$\nu$NS spectrum.} The CE$\nu$NS signal from reactor antineutrinos can be calculated as follows.
The differential CE$\nu$NS event rate with respect to the nuclear recoil energy $E_R$ is given by
\beq
\frac{dR}{d E_R} = N_T \int \frac{d\Phi}{dE_\nu}\frac{d\sigma}{dE_R} d E_\nu\,,
\label{eq:eventrate}
\eeq
where $N_T$ is the number of nuclei in the detector. 
The reactor antineutrino flux $\frac{d\Phi}{dE_\nu}$ is given by
\beq
\frac{d\Phi}{dE_\nu} = \frac{P}{4\pi d^2 \tilde{\epsilon} } \left(\frac{d N_\nu}{dE_\nu}\right)\,,
\eeq
where $P=2.96$~GW is the reactor thermal power, $d=10$~m is the distance between the reactor and detector, and $\tilde{\epsilon}=205.24$ MeV is the average energy released per fission.  We use the antineutrino spectrum per fission $\frac{d N_\nu}{d E_\nu}$ provided in Appendix A of Ref.~\cite{Aguilar-Arevalo:2019zme}.
The differential CE$\nu$NS cross section in the standard model (SM) is given by~\cite{Freedman:1973yd}
\begin{equation}
\label{eq:crossSM}
\frac{d\sigma_{SM}}{dE_R}=\frac{G_F^2M}{4\pi}q_W^2\left(1-\frac{ME_R}{2E_{\nu}^2}\right)F^2(\mathfrak{q})\,,
\end{equation}
where $M$ is the nuclear mass, $E_{\nu}$ is the antineutrino energy, $G_F$ is the Fermi coupling constant,  $q_W = N - (1 -4 \sin^2\theta_W)Z$ is the weak nuclear charge with $\theta_W$ the weak mixing angle, and $F(\mathfrak{q})$ is the Klein-Nystrand parameterization of the nuclear form factor as a function of the momentum transfer $\mathfrak{q}$~\cite{Klein:1999gv}.
The calculated signal is not sensitive to the specific choice of the commonly used form factors and its uncertainties because of the low momentum transfer in 
CE$\nu$NS with reactor antineutrinos~\cite{AristizabalSierra:2019zmy}.

%
\begin{figure*}[t]
	\centering
	\includegraphics[width=0.47\textwidth]{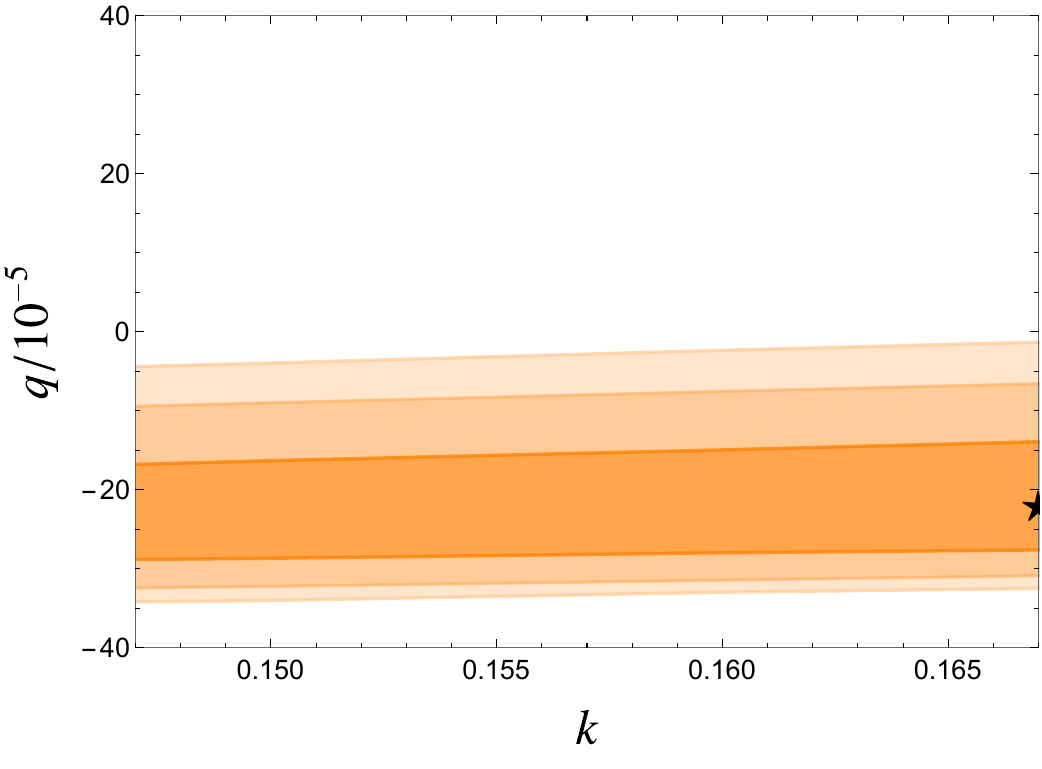}\,
	\includegraphics[width=0.47\textwidth]{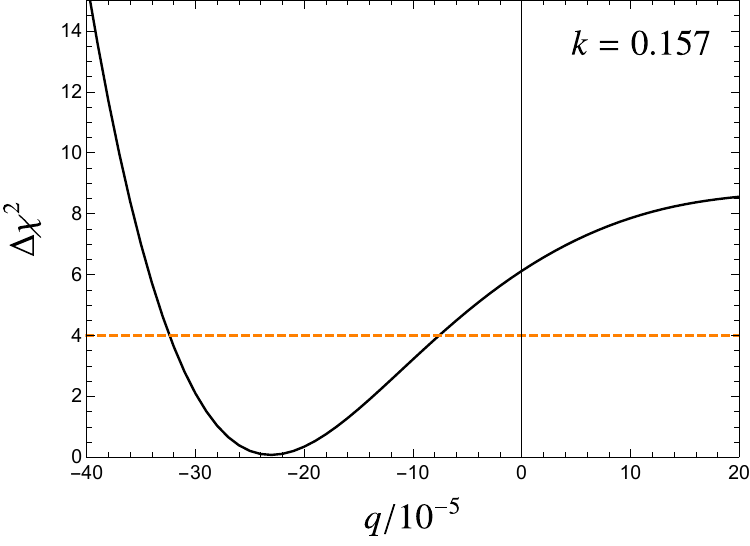}\,
	\caption{Left: the 1$\sigma$, 90\% CL, and $2\sigma$ allowed regions in ($k$, $q$) plane for the modified Lindhard model. The star marks the best fit point.
		Right: $\Delta \chi^2\equiv\chi^2(q)-\chi^2_{\rm min}$ for $k=0.157$.}
	\label{fig:kq}
\end{figure*}

However, the predicted CE$\nu$NS spectrum as a function of the {\it  measured} energy is strongly dependent on the germanium quenching factor. 
The quenching factor $Q$ is defined as the ratio of the observable recoil energy in a nuclear recoil $E_I$ (say in the form of ionization or scintillation) to the observable recoil energy in an electron recoil of the same total recoil energy $E_R$, i.e., $Q\equiv E_I/E_R$. The differential event rate with respect to ionization energy $E_I$ is
\beq
\frac{dR}{dE_I}= \frac{dR}{d E_R} \left(\frac{1}{Q}-\frac{E_I}{Q^2}\frac{dQ}{dE_I}\right) \,.
\eeq
 For $E_R\gtrsim 5\text{ keV}_{\rm nr}$, experimental measurements of the quenching factor are well described by the standard Lindhard model~\cite{Lindard}. 
Under the assumptions that the atomic binding energy of electrons is negligible, and the electronic stopping power is velocity proportional without a threshold velocity, the quenching factor for a recoiling nucleus with atomic number $Z$ 
is given by
\begin{align}
Q(E_R)&= \frac{k\,g(\epsilon)}{1+k\,g(\epsilon)}\,,
\label{eq:lindhard1}
\end{align}
where $g(\epsilon)$ is well fitted by~\cite{Lewin:1995rx}
\begin{align}
g(\epsilon) &= 3\,\epsilon^{0.15} + 0.7\,\epsilon^{0.6}+\epsilon\,,
\end{align}
with
$
 \epsilon = 11.5\,Z^{-\frac{7}{3}}\,E_R\,.
$
Here, $E_R$ is in keV, $\epsilon$ is a dimensionless parameter,  and $k$ is a measure of  the electronic energy loss. In the standard Lindhard model, 
$k \approx 0.157 $ for germanium.

For sub-keV nuclear recoils, the quenching factors are not well modeled due to uncertainties in nuclear scattering and stopping at low energies~\cite{Lindard2, Sorensen:2014sla}. A recent measurement of the germanium quenching factor obtained using multiple techniques shows a departure from the Lindhard model for nuclear recoil energies below 
$\sim1$~$\text{keV}_{\rm nr}$~\cite{Collar:2021fcl}; however, Ref.~\cite{Bonhomme:2022lcz} finds no discrepancy. These data can be explained by the Migdal effect~\cite{Migdal} in neutron scattering on Ge~\cite{Collar:2021fcl}, and the overall shape of the quenching factor can be parameterized by a modified Lindhard model~\cite{Sorensen:2014sla, Liao:2021yog},
\begin{align}
\label{mod}
Q(E_R)&= \frac{k\,g(\epsilon)}{1+k\,g(\epsilon)}-{q\over \epsilon}\,,
\end{align}
where the parameter $q$ is negative (positive) if the energy given to electrons is enhanced (cutoff).  
The atomic binding energy gives $q>0$ thereby explaining an anticipated cutoff in $Q$~\cite{Sorensen:2014sla}, while the Migdal effect modeled by $q<0$ leads to an enhancement at  low recoil energies~\cite{Liao:2021yog}.
Note that a nonzero $q$ mainly affects the quenching factor at low energies and leaves the high-energy behavior of the standard Lindhard model unchanged. 
Accounting for the energy resolution, the differential event rate with respect to the measured energy $E_M$ is~\cite{Aguilar-Arevalo:2019zme}
\beq
\frac{dR}{dE_M}= \frac{\int_{0}^{\infty} G(E_M,E_I,\sigma^2)\frac{dR}{dE_I} dE_I}{\int_{0}^{\infty} G(E_M,E_I,\sigma^2)dE_I}\,.
\label{eq:dRdEM}
\eeq
Here, we assume a Gaussian detector response, 
\beq
G(E_M,E_I,\sigma^2)=\frac{1}{\sqrt{2\pi\sigma^2}}\exp{\left[-\frac{(E_M-E_I)^2}{2\sigma^2}\right]}\,,
\eeq
where the energy resolution $\sigma$ is approximated by $
\sigma^2=\sigma_n^2+E_{I}\eta F$.
Here, $\sigma_n=68.5$~eV is the intrinsic electronic noise, $\eta=2.96$~eV is the average energy required for photons to create an electron-hole pair in germanium, and $F\approx0.105$ is the Fano factor taken from Ref.~\cite{datarelease}.

%

{\bf Quenching factor.}
We first reproduce the efficiency-corrected SM spectrum shown in Ref.~\cite{NCC}. The number of events with measured energy in the $i^{\rm th}$ bin $[E_M^i, E_M^{i+1}]$  is given by 
\beq
N_i=t\int_{E_M^i}^{E_M^{i+1}}\frac{dR}{d E_M} dE_M\,,
\label{eq:counts}
\eeq
where $t=289.2$~kg$\cdot$day is the exposure time, and the differential event rate $\frac{dR}{dE_M}$ is given by Eq.~(\ref{eq:dRdEM}). We assume that the high purity germanium isotope in the detector is $^{72}$Ge. 
We analyze the spectrum of residual counts after the best-fit background is subtracted; see Fig.~5 of Ref.~\cite{NCC}.

\begin{figure*}
	\includegraphics[width=0.47\textwidth]{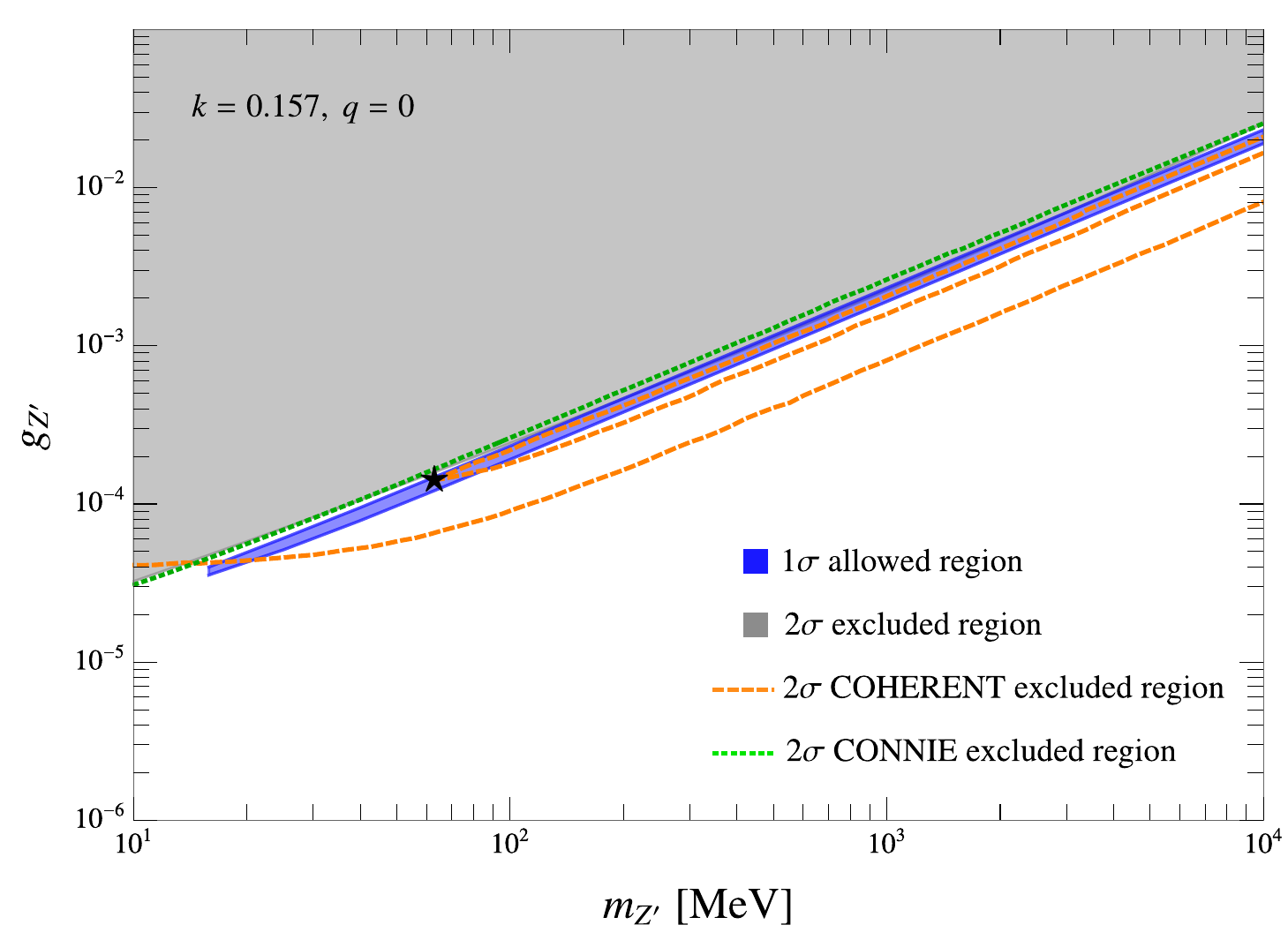}
	\includegraphics[width=0.47\textwidth]{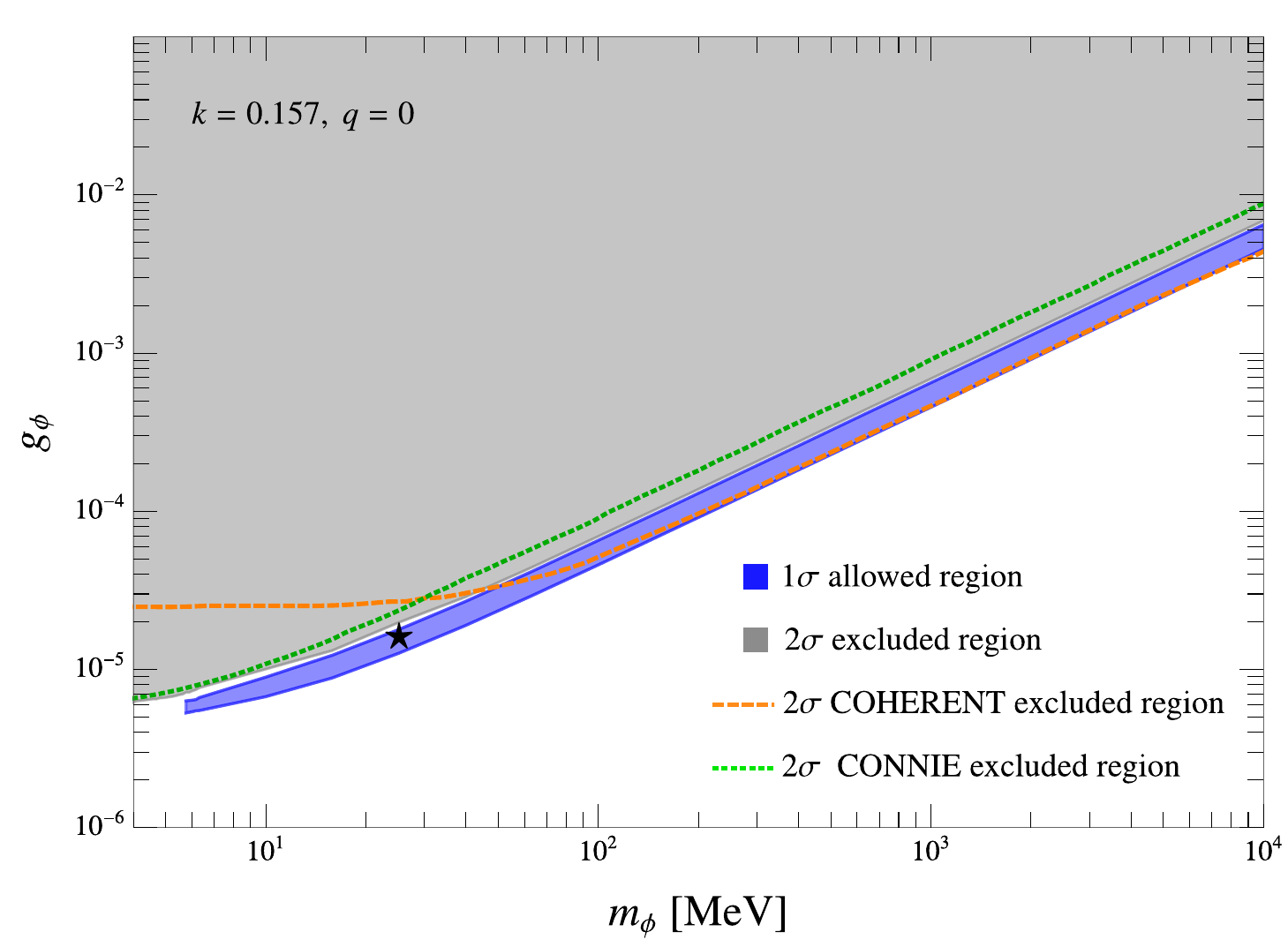}
	\includegraphics[width=0.47\textwidth]{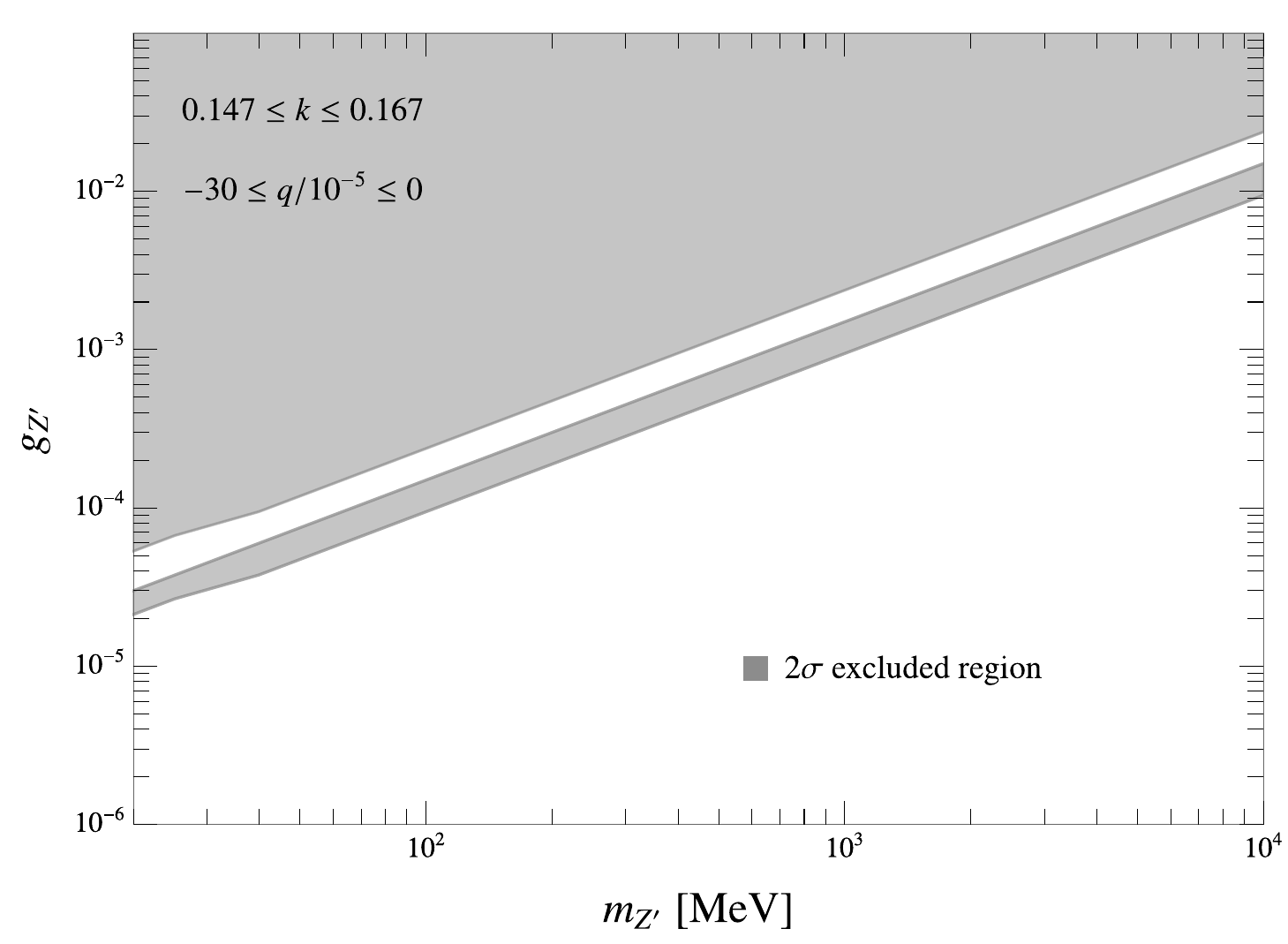}
	\includegraphics[width=0.47\textwidth]{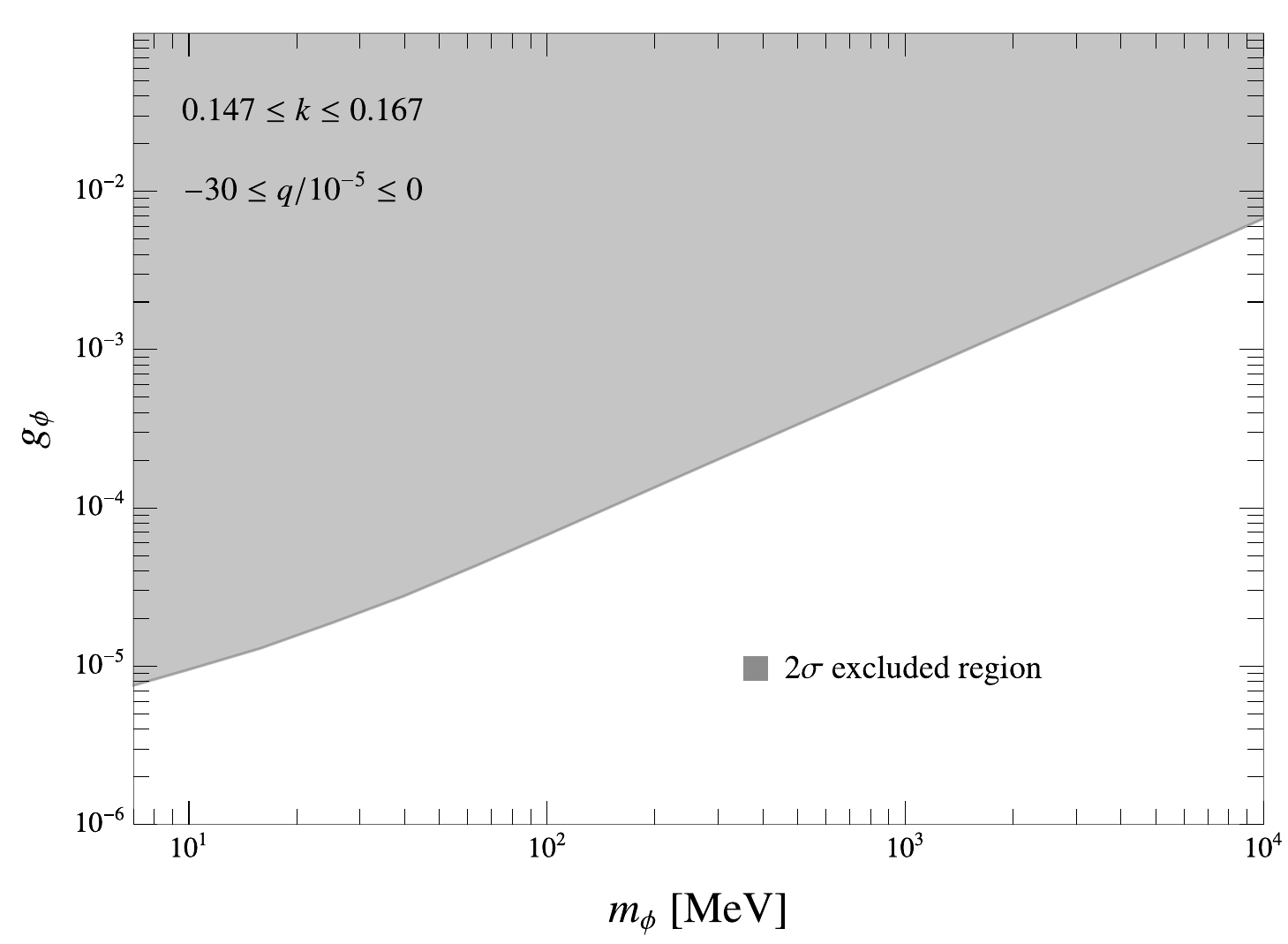}
	\caption{The 1$\sigma$ allowed region and 2$\sigma$ excluded region in the ($m_{Z^\prime}$, $g_{Z^\prime}$) [($m_{\phi}$, $g_\phi$)] plane for the light $Z^\prime$ [scalar] scenario. Upper panels: We assume $k=0.157$ and $q=0$ for the quenching factor. The best fit point in each scenario is marked by a star. The COHERENT and CONNIE excluded regions are taken from Ref.~\cite{Aguilar-Arevalo:2019zme},  with an appropriate rescaling of the $Z'$ coupling. 
Lower panels: We marginalize over quenching factor models with $0.147\leq k \leq 0.167$ and 
$-30\leq q/10^{-5} \leq 0$. No allowed region is shown because the SM is allowed within $1\sigma$.
	The smallest abscissae in the four panels  are different because we only show bounds for mediator masses for which less than half the expected events at NCC-1701 have $E_M > E_R$ (which is unphysical without the energy resolution function in Eq.~\ref{eq:dRdEM}).}
	\label{fig:contours}
\end{figure*}

We first study the implications of the measured CE$\nu$NS data for quenching factor models. 
To evaluate the statistical significance of a theoretical model, we define 
\begin{align}
\chi^2=\sum_i\left[\frac{N_\text{exp}^i-N_\text{th}^i(1+\alpha)}{\sigma_{i}}\right]^2+\left(\frac{\alpha}{\sigma_\alpha}\right)^2\,,
\end{align}
where $N^{i}_{{\rm exp}}$ is the measured number of residual counts per bin and $\sigma_i$ is the corresponding uncertainty, $N^{i}_{{\rm th}}$ is the expected number of events per bin calculated using Eq.~(\ref{eq:counts}), and $\sigma_\alpha=5\%$ is the percent uncertainty in the reactor neutrino flux normalization.  

We fit 20 bins in $E_M$ from 0.2 to 0.4 keV$_{\rm ee}$, and marginalize over the auxiliary parameter $\alpha$ to obtain $\chi^2_\text{min}$. 
For the SM with the standard Lindhard model for the quenching factor, 
we find $\chi^2_\text{min}=14.3$, which is a very good fit to the data. However, we consider the extent 
to which a modified Lindhard model improves the fit. We only consider $k$ values in the range of [0.147, 0.167], to be compatible with quenching factor measurements at high recoil energies~\cite{Lin:2007ka}. The 1$\sigma$, 90\% CL, and $2\sigma$ allowed regions in the ($k$, $q$) space are shown in Fig.~\ref{fig:kq}.  The best-fit point is located at $k=0.167$ and $q=-22.2\times 10^{-5}$, with $\chi^2_\text{min}=8.14$, which is a substantial improvement over the standard Lindhard model. This best-fit point is consistent with the direct quenching factor measurements of Ref.~\cite{Collar:2021fcl}, which can be parametrized by $(k,q<0)$, as shown in Ref.~\cite{Liao:2021yog}. Not surprisingly, NCC-1701 data provide an independent probe of the quenching factor. From the left panel of Fig.~\ref{fig:kq}, we see that the data are not sensitive to $k$, and prefer negative values of $q$. 
In the right panel, we plot  $\Delta \chi^2\equiv\chi^2(q)-\chi^2_\text{min}$ for $k=0.157$. Clearly, $q<0$ is preferred at $2.5\sigma$.

{\bf New physics.}
The measured CE$\nu$NS spectrum will be modified by new physics in the neutrino sector. We consider three simple new physics scenarios: 
(i) a light $Z^\prime$ that couples to neutrinos and quarks; (ii) a light scalar that couples to neutrinos and quarks; (iii) a large neutrino magnetic moment.

The differential cross section that includes contributions from the standard model (SM) and new universal flavor-conserving interactions mediated by a light vector $Z^\prime$ with mass $m_{Z^\prime}$ and coupling $g_{Z^\prime}$ is~\cite{review}
\begin{equation} 
\label{eq:crossV} 
\frac{d\sigma_{SM+Z^{\prime}}}{dE_R}=\left(1-\frac{q_{Z^{\prime}}}{q_W}\right)^2\frac{d\sigma_{SM}}{dE_R}\,,
\end{equation}
with the effective charge $q_{Z^{\prime}}$ given by~\footnote{Our convention for the $Z'$ coupling is related to that of Ref.~\cite{Liao:2017uzy} by $g_{Z^{\prime}} \equiv g/\sqrt{2}$.}
\begin{equation}
\label{eq:QLV}
q_{Z^{\prime}} = \frac{3\sqrt{2}\left(N+Z\right) g_{Z^{\prime}}^2}{G_F\left(2M E_R + m_{Z^{\prime}}^2\right)}\,.
\end{equation}
Equation~(\ref{eq:crossV}) shows that a light $Z^\prime$ can suppress the cross section via destructive interference. Also, for $m_{Z^\prime}\gg \sqrt{2M E_R}$, the scenario is degenerate with the SM if $q_{Z^{\prime}}=2 q_W$. This occurs for
 \begin{align}
 \frac{g_{Z^{\prime}}}{m_{Z^\prime}}=\sqrt{\frac{\sqrt{2}G_F\left[N - (1 -4 \sin^2\theta_W)Z\right]}{3(N+Z)}}\,.
 \label{eq:degeneracy}
 \end{align}
 
The differential cross section that includes new universal flavor-conserving interactions mediated by a light scalar $\phi$ with mass
$m_\phi$ and coupling $g_\phi$
is~\cite{review}
\begin{equation}
\label{eq:crossS}
\frac{d\sigma_{SM+\phi}}{dE_R}=~\frac{d\sigma_{SM}}{dE_R}+ \frac{d\sigma_{\phi}}{dE_R}\,,
\end{equation}
where
\begin{equation}
\frac{d\sigma_{\phi}}{dE_R}~=~\frac{G_F^2}{4\pi}q_{\phi}^2\frac{2ME_R}{E_{\nu}^2}MF^2(\mathfrak{q})\,,
\end{equation}
with $q_{\phi}$ given by
\begin{equation}
\label{eq:QLS}
q_{\phi}~=~\frac{\left(14 N + 15.1 Z\right)g_{\phi}^2}{\sqrt{2}G_F\big(2ME_R + m_{\phi}^2\big)}\,.
\end{equation}

The differential cross section that includes a large flavor-universal neutrino magnetic moment 
$\mu_\nu$
is~\cite{review}
\begin{equation}
\label{eq:crossmu}
\frac{d\sigma_{SM+\mu_\nu}}{dE_R}=~\frac{d\sigma_{SM}}{dE_R}+ \frac{\pi\alpha^2Z^2F^2(\mathfrak{q})}{m_e^2}\left(\frac{1}{E_R}-\frac{1}{E_\nu}\right)\left(\frac{\mu_\nu}{\mu_B}\right)^2\,,
\end{equation}
where $m_e$ is the electron mass and $\mu_B$ is the Bohr magneton.


%
\begingroup
\setlength{\tabcolsep}{10pt} 
\renewcommand{\arraystretch}{1.} 
\begin{table*}[t]
	\centering
	\begin{tabular}{c |  c|  c|  c|  c}
		\toprule
		scenarios   & $k$  & $q/10^{-5}$ & model parameters & $\chi^2_\text{min}$/d.o.f.\\
		\hline
		\midrule
		SM  w/ standard Lindhard& $0.157$ & $0$ & - & 14.34/19\\
		\midrule
		SM w/ modifed Lindhard w/ fixed $k$ & $0.157$ & $-23.8$ & - & 8.28/18\\
		\midrule
		SM w/ modified Lindhard w/ $0.147\le k\le 0.167$ & $0.167$ & $-22.2$ & - & 8.14/17\\
		\midrule
		light $Z^{\prime}$  & $0.157$ & $0$ & $m_{Z^\prime}=63.1$ MeV, $g_{Z^\prime}=1.4\times10^{-4}$ & 9.09/17\\ 
		\midrule
		light scalar  & $0.157$ & $0$ & $m_{\phi}=25.1$ MeV, $g_{\phi}=1.6\times10^{-5}$ & 7.77/17\\
		\midrule
		neutrino magnetic moment  & $0.157$ & $0$ & $\mu_\nu=2.5\times10^{-10} \mu_B$ & 11.71/18\\
		\toprule
	\end{tabular}
	\caption{Values of $\chi^2_{\rm min}$/dof for the SM with the standard/modified Lindhard model, and for the new physics scenarios with the standard Lindhard model. }
	\label{Table: chi2}
\end{table*}
\endgroup

\begin{figure}[t]
	\centering
	\includegraphics[width=0.47\textwidth]{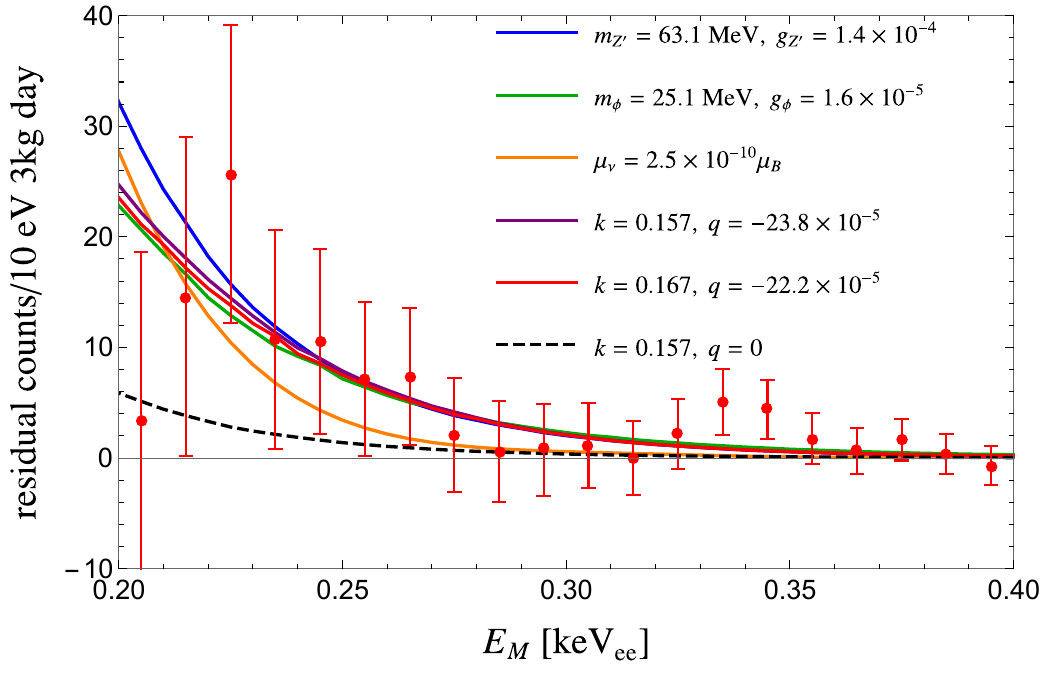}\,
	\caption{The spectra for the points in Table~\ref{Table: chi2}. The data points are the residual counts after  subtraction of the best-fit background.}
	\label{fig:spectrum}
\end{figure}

To place constraints on the  new physics scenarios, we scan over possible values of the coupling and mediator mass in the light $Z^\prime$ and scalar cases, and consider two different treatments of the quenching factor: (i) we assume the standard Lindhard model is valid, and (ii) we marginalize over the two parameters of the modified Lindhard model to reduce the dependence on the quenching factor model. 
The 1$\sigma$ allowed regions and $2\sigma$ excluded regions are shown in Fig.~\ref{fig:contours}, and the best fit points and $\chi^2_{\rm min}$/dof values are listed in Table~\ref{Table: chi2}. The best-fit spectra are shown in Fig.~\ref{fig:spectrum}.
The $1\sigma$ allowed regions in the upper panels of Fig.~\ref{fig:contours} (with $k=0.157$ and $q=0$) are very narrow for both the $Z^\prime$ and scalar cases. 
 The allowed region above $m_{Z'}=60$~MeV is also allowed  by COHERENT data because of the degeneracy in Eq.~(\ref{eq:degeneracy}). 
We see that the data show a mild preference for the new physics scenarios compared to the SM if the standard Lindhard model is assumed for the quenching factor. In the lower panels, 
we marginalize over the quenching factor with $0.147\leq k \leq 0.167$ and 
$-30\leq q/10^{-5} \leq 0$. Only $2\sigma$ excluded regions are shown because the SM is allowed within $1\sigma$. The excluded region in the $Z^\prime$ case is split into two parts as a result of the degeneracy. We conclude that the constraints are qualitatively affected by the quenching factor model, essentially through its dependence on $q$. 
Note that energy resolution effects permit events with $E_R < E_M$.  To avoid the circumstance of too many such events, we require that at least half the expected events have $E_R > E_M$. 
We only show bounds for mediator masses that meet this criterion, which explains why the bounds do not flatten out as the mediator mass decreases.

The best-fit value of the neutrino magnetic moment, and the corresponding $\chi^2_{\rm min}$ 
are provided in Table~\ref{Table: chi2}.
The 90\% CL bound from NCC-1701 is $\mu_\nu<4.0\times10^{-10} \mu_B$, which is an order of magnitude weaker than the current 90\% CL  bound on the electron neutrino magnetic moment, $\mu_\nu<2.9\times 10^{-11} \mu_B$~\cite{Beda:2012zz}.


%

{\bf Summary.} 
The first evidence for CE$\nu$NS using reactor antineutrinos is consistent with the SM. 
However, we find that the standard Lindhard model with $k=0.157$ and $q=0$ is ruled out at $2.5\sigma$. A negative value of $q$ is preferred by the NCC-1701 data at
2$\sigma$. This may be related to the Migdal effect in neutron scattering on germanium.
The low energies of reactor neutrinos enable us to place stringent bounds  on new vector and scalar mediators that couple to neutrinos and quarks.  However, these bounds are clearly dependent on how the quenching factor is modeled at low recoil energies.
If the standard Lindhard model correctly describes the quenching factor, the data may indicate a light vector or scalar mediator, or a large neutrino magnetic moment.
We await more data.

 {\it Acknowledgments.} 
We thank J.~Collar for providing us with a wee note on NCC-1701 data and for helpful discussions.
J.L. is supported by the National Natural Science Foundation of China under Grant No. 11905299 and Guangdong Basic and Applied Basic Research Foundation under Grant No. 2020A1515011479. H.L. is supported by ISF, BSF and Azrieli foundation. D.M. is supported in part by the U.S. Department of Energy under Grant No. de-sc0010504.



\begin{thebibliography}{99}
\bibitem{Freedman:1973yd} 
D.~Z.~Freedman,
Phys.\ Rev.\ D {\bf 9}, 1389 (1974).

\bibitem{Akimov:2017ade}
D.~Akimov \textit{et al.} [COHERENT],
Science \textbf{357}, no.6356, 1123-1126 (2017)
[arXiv:1708.01294 [nucl-ex]].


\bibitem{Akimov:2020pdx}
D.~Akimov \textit{et al.} [COHERENT],
Phys. Rev. Lett. \textbf{126}, no.1, 012002 (2021)
[arXiv:2003.10630 [nucl-ex]].

\bibitem{review}
For a review, see D.~Papoulias, T.~Kosmas and Y.~Kuno,
Front. in Phys. \textbf{7}, 191 (2019)
[arXiv:1911.00916 [hep-ph]].

\bibitem{Aguilar-Arevalo:2019jlr}
A.~Aguilar-Arevalo \textit{et al.} [CONNIE],
Phys. Rev. D \textbf{100}, no.9, 092005 (2019)
[arXiv:1906.02200 [physics.ins-det]].

\bibitem{Bonet:2020awv}
H.~Bonet \textit{et al.} [CONUS],
Phys. Rev. Lett. \textbf{126}, no.4, 041804 (2021)
[arXiv:2011.00210 [hep-ex]].

\bibitem{Colaresi:2021kus}
J.~Colaresi, J.~I.~Collar, T.~W.~Hossbach, A.~R.~L.~Kavner, C.~M.~Lewis, A.~E.~Robinson and K.~M.~Yocum,
Phys. Rev. D \textbf{104}, no.7, 072003 (2021)
[arXiv:2108.02880 [hep-ex]].

\bibitem{NCC}
J.~Colaresi, J.~I.~Collar, T.~W.~Hossbach, C.~M.~Lewis and K.~M.~Yocum,
[arXiv:2202.09672 [hep-ex].]

\bibitem{Collar:2021fcl}
J.~I.~Collar, A.~R.~L.~Kavner and C.~M.~Lewis,
Phys. Rev. D \textbf{103}, no.12, 122003 (2021)
[arXiv:2102.10089 [nucl-ex]].

\bibitem{Aguilar-Arevalo:2019zme}
A.~Aguilar-Arevalo \textit{et al.} [CONNIE],
JHEP \textbf{04}, 054 (2020)
[arXiv:1910.04951 [hep-ex]].

\bibitem{Klein:1999gv}
S.~R.~Klein and J.~Nystrand,
Phys. Rev. Lett. \textbf{84}, 2330-2333 (2000)
[arXiv:hep-ph/9909237 [hep-ph]].

\bibitem{AristizabalSierra:2019zmy}
D.~Aristizabal Sierra, J.~Liao and D.~Marfatia,
JHEP \textbf{06}, 141 (2019)
[arXiv:1902.07398 [hep-ph]].


\bibitem{Lindard}
J.~Lindhard, V.~Nielsen, M.~Scharff, and P.~Thomsen, Kgl. Danske Videnskab., Selskab. Mat. Fys. Medd. 33,
10 (1963).

\bibitem{Lewin:1995rx}
J.~D.~Lewin and P.~F.~Smith,
Astropart. Phys. \textbf{6}, 87-112 (1996).


\bibitem{Lindard2}
J. Lindhard, M. Scharff and H.E. Schiott, Mat. Fys. Medd. Dan. Vid. Selsk. 33 14 (1963).



\bibitem{Sorensen:2014sla}
P.~Sorensen,
Phys. Rev. D \textbf{91}, no.8, 083509 (2015)
[arXiv:1412.3028 [astro-ph.IM]].

\bibitem{Bonhomme:2022lcz}
A.~Bonhomme, H.~Bonet, C.~Buck, J.~Hakenm\"uller, G.~Heusser, T.~Hugle, M.~Lindner, W.~Maneschg, R.~Nolte, T.~Rink, E. Pirovano and H. Strecker,
[arXiv:2202.03754 [physics.ins-det]].

\bibitem{Migdal}
A. Migdal, J.Phys. (USSR) 4, 449  (1941).


\bibitem{Liao:2021yog}
J.~Liao, H.~Liu and D.~Marfatia,
Phys. Rev. D \textbf{104}, no.1, 015005 (2021)
[arXiv:2104.01811 [hep-ph]].


\bibitem{datarelease}
J. Collar, NCC-1701 data release.

\bibitem{Lin:2007ka}
S.~T.~Lin \textit{et al.} [TEXONO],
Phys. Rev. D \textbf{79}, 061101 (2009)
[arXiv:0712.1645 [hep-ex]].


\bibitem{Liao:2017uzy}
J.~Liao and D.~Marfatia,
Phys. Lett. B \textbf{775}, 54-57 (2017)
[arXiv:1708.04255 [hep-ph]].


\bibitem{Beda:2012zz}
A.~Beda, V.~Brudanin, V.~Egorov, D.~Medvedev, V.~Pogosov, M.~Shirchenko and A.~Starostin,
Adv. High Energy Phys. \textbf{2012}, 350150 (2012).







\end{thebibliography}
\end{document}